\documentclass[superscriptaddress, preprint, amsmath, amssymb, aps, physrev]{revtex4-2}

\usepackage{float}
\usepackage[english]{babel} 
\usepackage{subcaption} % This package is used for subfigures
\DeclareUnicodeCharacter{2212}{-}
\usepackage{booktabs}  % For professional looking tables
\usepackage{siunitx}   % For proper number alignment
\usepackage{amsmath}
\usepackage{chemformula}
\usepackage{mhchem}
\usepackage[utf8]{inputenc}

\usepackage{multirow}

\usepackage{chemformula}

\begin{document}
\title{ Universal Stability of Ga Split Vacancies across $\alpha$-, $\beta$-, and $\kappa$-\ce{Ga2O3} Polymorphs: A Machine-Learning Accelerated Study}
\author{Mohamed Abdelilah Fadla}
\email{m.fadla@qub.ac.uk}
\affiliation{School of Mathematics and Physics, Queen's University Belfast, University Road, Belfast BT7 1NN, UK}
\author{Myrta Gr\"{u}ning}
\affiliation{School of Mathematics and Physics, Queen's University Belfast, University Road, Belfast BT7 1NN, UK}
\affiliation{European Theoretical Spectroscopy Facility}
\author{Lorenzo Stella}
\affiliation{School of Mathematics and Physics, Queen's University Belfast, University Road, Belfast BT7 1NN, UK}

\begin{abstract}

Split Ga vacancies are the dominant native acceptor in $\beta$-\ce{Ga2O3}; however, their role in $\alpha$ and $\kappa$ phases has been largely overlooked or assumed to be unfavorable. A detailed understanding of these defects is critical for tailoring the electrical conductivity and optical properties and optimising \ce{Ga2O3}-based devices. In this work, we used machine learning interatomic potentials (MLIPs) to accelerate the discovery of non-local defect reconstructions, followed by HSE06 hybrid DFT to accurately quantify defect properties of single vacancy $V_{\text{Ga}}$, split vacancy $V_{\text{Ga}}^{\text{i}}$ and substitutional donors ($\mathrm{Hf_{Ga}}$ and $\mathrm{Si_{Ga}}$) across a wide range of experimentally relevant conditions for the oxygen chemical potential.
We find that split vacancies are the ground-state vacancy for all studied polymorphs ($\beta$, $\alpha$, and $\kappa$). Split vacancies are more stable than simple vacancies by ~0.75 eV ($\beta$), ~0.41 eV ($\alpha$), and ~0.14 eV ($\kappa$). Notably, MLIPs correctly identified the specific split-vacancy ground states and yielded an energetic ordering of symmetry-inequivalent defect configurations in excellent agreement with HSE06 results. While Hf and Si show low formation energy and act as shallow donors, especially under oxygen-poor conditions, their efficiency is limited by split-vacancy compensation. The growth under oxygen-poor conditions is a universal requirement to suppress these defects and achieve high n-type conductivity across the \ce{Ga2O3} polymorph.
\end{abstract}

\maketitle
\newpage
\section{Introduction}

In recent years, $\beta$-\ce{Ga2O3} has attracted great attention  due to its outstanding properties, including a wide band gap of ($\sim$4.7~eV),  high breakdown electric field ($\sim$8~MV/cm) and wide-range tunability of carrier concentration ($10^{15}$ to $10^{20}$ cm$^{-3}$). \cite{varley2022wide,galazka_bulk_2014,tadjer2022toward,higashiwaki2017state,pearton2018review} These characteristics, combined with the economical advantage of the widespread availability of melt-grown large-size single crystals, have accelerated the development of unipolar high-voltage field-effect transistors and Schottky barrier diodes. \cite{pearton2018review, higashiwaki2016current}

More recently, other polymorphs such as $\alpha$- and $\kappa$-\ce{Ga2O3} have gained increased interest due to their potentially superior properties.\cite{safieddine2022comparative,kobayashi2019energetics,wheeler2020phase,roberts2019low,nicol2023hydrogen,ma2019heteroepitaxial,seacat2020orthorhombic,maccioni2016phase,kim2018first} Compared to $\beta$-\ce{Ga2O3}, the most thermodynamically stable phase, $\alpha$-\ce{Ga2O3}, with a trigonal crystal structure ($\ce{R\bar{3}c}$),  has a wider band gap ($\sim$5.3~eV), higher breakdown field (9.5~MV/cm) and better thermal conductivity. \cite{kobayashi2019energetics,wheeler2020phase,roberts2019low,nicol2023hydrogen} 
$\alpha$-\ce{Ga2O3} has the same crystal structure as sapphire ($\alpha$-\ce{Al2O3}) and high-quality heteroepitaxial thin films of $\alpha$-\ce{Ga2O3}  can be grown on a relatively low-cost substrate. \cite{shinohara2008heteroepitaxy,ma2019heteroepitaxial}  Reported fabrication of $\alpha$-\ce{Ga2O3}-based devices includes metal-semiconductor field-effect transistors and Schottky barrier diodes. \cite{kaneko2018power,dang2015mist} The orthorhombic Pna2$_1$ structure $\kappa$-\ce{Ga2O3} (referred to in earlier studies as $\epsilon$,\cite{cora2017real,seacat2020orthorhombic}) exhibits a large bandgap of ($\sim$5~eV) and interestingly may possess ferroelectric properties,\cite{maccioni2016phase,kim2018first,ranga2020highly,seacat2020orthorhombic} resulting in large spontaneous electrical polarizations. \cite{maccioni2016phase,kim2018first,cho2018epitaxial,ranga2020highly} 

Understanding native defects in \ce{Ga2O3} is critical for controlling electrical conductivity, optical properties, and device performance. In $\beta$-\ce{Ga2O3}, the Ga split-vacancy dominates as the primary native acceptor.\cite{varley2011hydrogenated}  Previous reports investigated its role as a deep acceptor in reducing n-type dopability and increasing carrier compensation. \cite{varley2011hydrogenated,karjalainen2021split,tuomisto2023ga} Theoretical calculations and experimental measurements, including positron annihilation spectroscopy and transmission electron microscopy, indicate the presence of Ga split vacancies with high concentrations in $\beta$-\ce{Ga2O3}.\cite{karjalainen2021split,tuomisto2023ga} 

Despite these promising properties across \ce{Ga2O3} polymorphs, the behavior of split-vacancy defects in the $\alpha$- and $\kappa$-phases remains largely unexplored or assumed to be unfavorable\cite{polyakov2022point} with few studies that focused on the fundamental properties of $\alpha$- and $\kappa$-\ce{Ga2O3} polymorphs.\cite{seacat2020orthorhombic,mu2022phase} Given the critical impact of these defects on carrier compensation in $\beta$-\ce{Ga2O3}, a comprehensive investigation of defect energetics and ionisation levels across these polymorphs is required.

In this work, we employed a machine learning interatomic potentials (MLIPs) to accelerate defect discovery, followed by first-principles investigation using a hybrid functional to investigate split Ga vacancy and donor defects ($\mathrm{Si_{Ga}}$ and $\mathrm{Hf_{Ga}}$) across  $\beta$, $\alpha$ and $\kappa$ \ce{Ga2O3} polymorphs. We studied defect formation energies under oxygen-rich and oxygen-poor growth conditions and found that the split Ga vacancy is the ground-state configuration in all three polymorphs.

\section{Methodology}

All DFT calculations were performed using projector augmented wave (PAW) method as implemented in the Vienna ab initio simulation package (VASP) \cite{kresse_efficient_1996,kresse1996efficiency,kresse1993ab, blochl_projector_1994}. For accurate structural and electronic properties, Heyd-Scuseria-Ernzerhof hybrid functional (HSE06) with 32\% Hartree-Fock exact exchange \cite{10.1063/1.1564060,krukau_influence_2006} was employed. All defective structures were fully optimised using a convergence criterion of \SI{0.01}{\electronvolt\per\angstrom} for the largest force components and an energy tolerance set at \SI{1e-6}{\electronvolt}. For bulk primitive cells of three polymorphs, sequential full optimisation steps were performed until the forces converged within a single ionic step. 

All defect calculations were performed in a 120-atom supercell, with a $k$-point mesh set to $2 \times 2 \times 2$ and a cutoff energy of \SI{600}{\electronvolt}. Tetrahedron smearing \cite{blochl1994improved}  was employed for accurate density of states (DOS). Spin polarisation was taken into account for defect charge with odd numbers of electrons.

We extensively used \texttt{doped} \cite{Kavanagh2024} defect simulation package for defect structure generation, parsing the calculations, and analysing the results. For initial screening of defect configurations, electrostatic energies were calculated using a basic Ewald summation model with fully-ionised charge states and a compensating background charge, which is implemented in \texttt{pymatgen} and \texttt{EwaldSummation} tools. \cite{kavanagh_identifying_2025,ong2013python,toukmaji1996ewald} 
MLIPs, including MACE foundation models and a multi-head fine-tuned variant trained on HSE06 data,\cite{batatia2022mace,batatia2025foundationmodelatomisticmaterials} were used for efficient screening and pre-relaxation of the selected single and split vacancy configurations(further details on model benchmarking and fine-tuning are provided in the SI).

The defect formation energies, $E_{\mathrm{f}}\left[X^{q}\right]$, are computed using the standard approach \cite{van_de_walle_first-principles_2004}, as 
\begin{equation}
E_{\mathrm{f}}\left[X^{q}\right]=E_{\text {tot }}\left[X^{q}\right]-E_{\text {tot }}[\text { bulk }]-\sum_i n_{i} \mu_{i}+q (E_{\mathrm{F}} + E_{\text{VBM}}) + E_{\text {corr }}^q,
\label{eq:1}
\end{equation}
where $E_{\text{tot}}[X^q]$ is the total energy of the defective supercell with charge $q$, $E_{\text{tot}}[\text{bulk}]$ is the total energy of the same size pristine supercell, $n_i$ is the number of i atoms added or removed from the defective supercell, $\mu_i$ is the chemical potential relative to the elemental phase, and $E_{\text{VBM}}$ is the valence band maximum and $E_{\mathrm{F}}$ is the Fermi level referenced to the VBM. The finite-size effect correction, $E_{\text {corr}}$, is defined according to the extended Freysoldt, Neugebauer, and Van de Walle (eFNV) scheme of Kumagai and Oba. \cite{kumagai_electrostatics-based_2014,freysoldt_fully_2009} 

For the $\beta$-\ce{Ga2O3} phase, a  $1 \times 3 \times 2$ expansion of C2/m conventional unit cell was used, as in our previous work.\cite{fadla2025tailoring} For the other polymorphs, we employed the scheme implemented in  \texttt{doped} \cite{Kavanagh2024} package, which uses an efficient algorithm to compute minimum image distances for supercell choice optimisation (transformation matrices are given in the Supplementary Information). All these supercells were tested for converged formation energy and flat electrostatic site potential at the sampling region, far from the defect location. For interstitial sites, candidate sites were selected using Voronoi tessellation,\cite{Kavanagh2024} considering multiplicities and symmetry-equivalent coordinates. The \texttt{Sumo} package was used for band structure pre and post-processing, including k-point generation, effective mass calculation and band structure plotting. \cite{ganose2018sumo}

\newpage
\section{Results and Discussions}
\subsection{Bulk properties of \ce{Ga2O3} polymorphs}

The monoclinic phase~($\beta$-\ce{Ga2O3}), the most extensively studied structure, has two inequivalent Ga sites with octahedral and tetrahedral coordination. 
The rhombohedral~($\alpha$-\ce{Ga2O3}) phase, similar to the Al$_2$O$_3$ corundum structure, contains a single octahedrally coordinated Ga site. 
The orthorhombic phase~($\kappa$-\ce{Ga2O3}), a polar structure, includes four distinct Ga sites exhibiting tetrahedral and octahedral coordination. 
Two octahedral sites exhibit large octahedral distortions with a single longer bond---were alternatively described as pentahedral coordination (Figure~S1 depicts supercell structures of \ce{Ga2O3} polymorphs).  Table~1 lists the HSE06-calculated equilibrium lattice parameters, band gaps, and effective masses.
\begin{table}[htbp]
\centering
\caption{Calculated (HSE06 level) and available previous experimental lattice parameters, band gaps, formation energies, and effective masses for the \ce{Ga2O3} polymorphs.}
\label{tab:Ga2O3_properties}
\begin{tabular}{lcccccc}
\toprule
& \multicolumn{2}{c}{$\alpha$-\ce{Ga2O3}} & \multicolumn{2}{c}{$\beta$-\ce{Ga2O3}} & \multicolumn{2}{c}{$\kappa$-\ce{Ga2O3}} \\
\cmidrule(lr){2-3} \cmidrule(lr){4-5} \cmidrule(lr){6-7}
Property & Calc. & Expt. & Calc. & Expt. & Calc. & Expt. \\
\midrule
Band gap (eV) & 5.283 & 5.3 \cite{ahmadi2019materials} & 4.825 & 4.8 \cite{ahmadi2019materials} & 4.935 & 4.91  \cite{kneiss2019tin} \\
Electron effective mass & 0.274--0.281 & -- & 0.275--0.284 & -- & 0.275--0.280 & -- \\

\midrule
\multirow{4}{*}{Lattice parameters (\si{Å})} 
& a = 4.97 & a = 4.983 \cite{marezio1967bond} & a = 12.23 & a = 12.21 \cite{aahman1996reinvestigation} & a = 5.029 & a = 5.0463 \cite{cora2017real}\\
& c = 13.41 & c = 13.433 \cite{marezio1967bond} & b = 3.034 & b = 3.04 \cite{aahman1996reinvestigation} & b = 8.649 & b = 8.7021\cite{cora2017real} \\
& -- & -- & c = 5.792 & c = 5.80 \cite{aahman1996reinvestigation} & c = 9.266 & c = 9.2833\cite{cora2017real} \\
& -- & -- & $\beta = \SI{103.83}{\degree}$ & $\beta = \SI{103.83}{\degree}$ \cite{aahman1996reinvestigation} & -- & -- \\
\bottomrule
\end{tabular}
\end{table}

The obtained structural and band gap values show excellent agreement with experiments, confirming the accuracy and predictivity of the computational approach. 
Among the \ce{Ga2O3} polymorphs, $\alpha$-\ce{Ga2O3} displays the largest band gap, 5.28~eV (5.3~eV expt.), which correlates with its slightly higher breakdown voltage of 9.5~MV/cm. 
All compounds exhibit a similar low electron effective mass of $\sim 0.28\, m_0$ and heavy holes due to their flat valence band maximum (see band structure in Figures~S2).

\subsection{Identification and stability of Ga split vacancies}

Ga single vacancies (V$_{\text{Ga}}$) can be directly created by removing a Ga atom, and their stability is assessed using defect formation energies.  For split-vacancy generation, we followed the same approach recently used in Ref.~\cite{kavanagh_identifying_2025}, to generate symmetry-inequivalent configurations for (V$_{\text{Ga}}$--Ga$_{\text{i}}$--V$_{\text{Ga}}$) where Ga$_{\text{i}}$--V$_{\text{Ga}}$ distances are less than 5~\si{\angstrom}.

We first evaluated electrostatic energy differences to identify the most favourable structures and exclude possible higher-energy configurations. We note that electrostatic screening was successfully used to reduce the search space and accelerate the identification of lower-energy split vacancies in similar compounds. \cite{kavanagh_identifying_2025} Next, we used MLIPs for rapid screening and pre-relaxation of selected configurations (more details on model evaluation are given in the SI).  Subsequently, we performed HSE06 level relaxations for the lowest-energy configurations and all single-vacancy configurations (\emph{e.g.}, four for the $\kappa$-\ce{Ga2O3} phase). 

Fully ionized +3 defect configurations were used to effectively enable electrostatic screening---one of the main contributors to the formation energy is electrostatic effects.  Additionally, V$_{\text{Ga}}^{-3}$ has the lowest formation energy across a wide Fermi-level range, particularly near the conduction band minimum, which is highly relevant for investigating compensation mechanisms in n-type \ce{Ga2O3}.  Table 2 reports the calculated relative energies of split and single vacancies with respect to the lowest-energy configurations.

\begin{table}[h]
\centering
\caption{
Relative total energies in $\beta$-\ce{Ga2O3}, $\alpha$-\ce{Ga2O3} and $\kappa$-\ce{Ga2O3}  (eV) of fully ionised gallium single and split vacancy defects, computed using MP-0b3 foundation model, fine-tuned MACE and HSE06 functional. Energies are referenced to the lowest-energy single vacancy ($V_{\text{Ga}}$)  in each polymorph. Additional models are provided in the Supplementary Information.}
\label{tab:defect-energies}
\begin{tabular}{llccc}
\hline
\textbf{Polymorph} & \textbf{Defect} & \textbf{MP-0b3} & \textbf{Fine-tuned MACE} & \textbf{HSE06} \\
\hline
\multirow{5}{*}{$\beta$-\ce{Ga2O3}}
 & $V_{\text{Ga}}^{\text{ic}}$ & -0.71 & \textbf{-0.69} & -0.75 \\
 & $V_{\text{Ga}}^{\text{ib}}$ & -0.45 & \textbf{-0.29} & -0.20 \\
 & $V_{\text{Ga}}^{\text{ia}}$ & 0.04  & \textbf{0.13}  & 0.12  \\
 & $V_{\text{Ga}_2}$          & 0.00  & \textbf{0.00}  & 0.00  \\
 & $V_{\text{Ga}_1}$          & 0.15  & \textbf{0.30}  & 0.51  \\
\hline
\multirow{2}{*}{$\alpha$-\ce{Ga2O3}}
 & $V_{\text{Ga}}^{\text{i}}$ & -0.44 & \textbf{-0.31} & -0.41 \\
 & $V_{\text{Ga}}$            & 0.00  & \textbf{0.00}  & 0.00  \\
\hline
\multirow{5}{*}{$\kappa$-\ce{Ga2O3}}
 & $V_{\text{Ga}}^{\text{i}}$ & -0.11 & \textbf{-0.02} & -0.14 \\
 & $V_{\text{Ga}_2}$          & 0.00  & \textbf{0.00}  & 0.00  \\
 & $V_{\text{Ga}_1}$          & 0.17  & \textbf{0.40}  & 0.47  \\
 & $V_{\text{Ga}_3}$          & 0.82  & \textbf{1.02}  & 1.09  \\
 & $V_{\text{Ga}_4}$          & 1.08  & \textbf{1.25}  & 1.43  \\
\hline
\end{tabular}
\end{table}

\begin{figure}[h]
     \centering
\includegraphics[width=7cm]{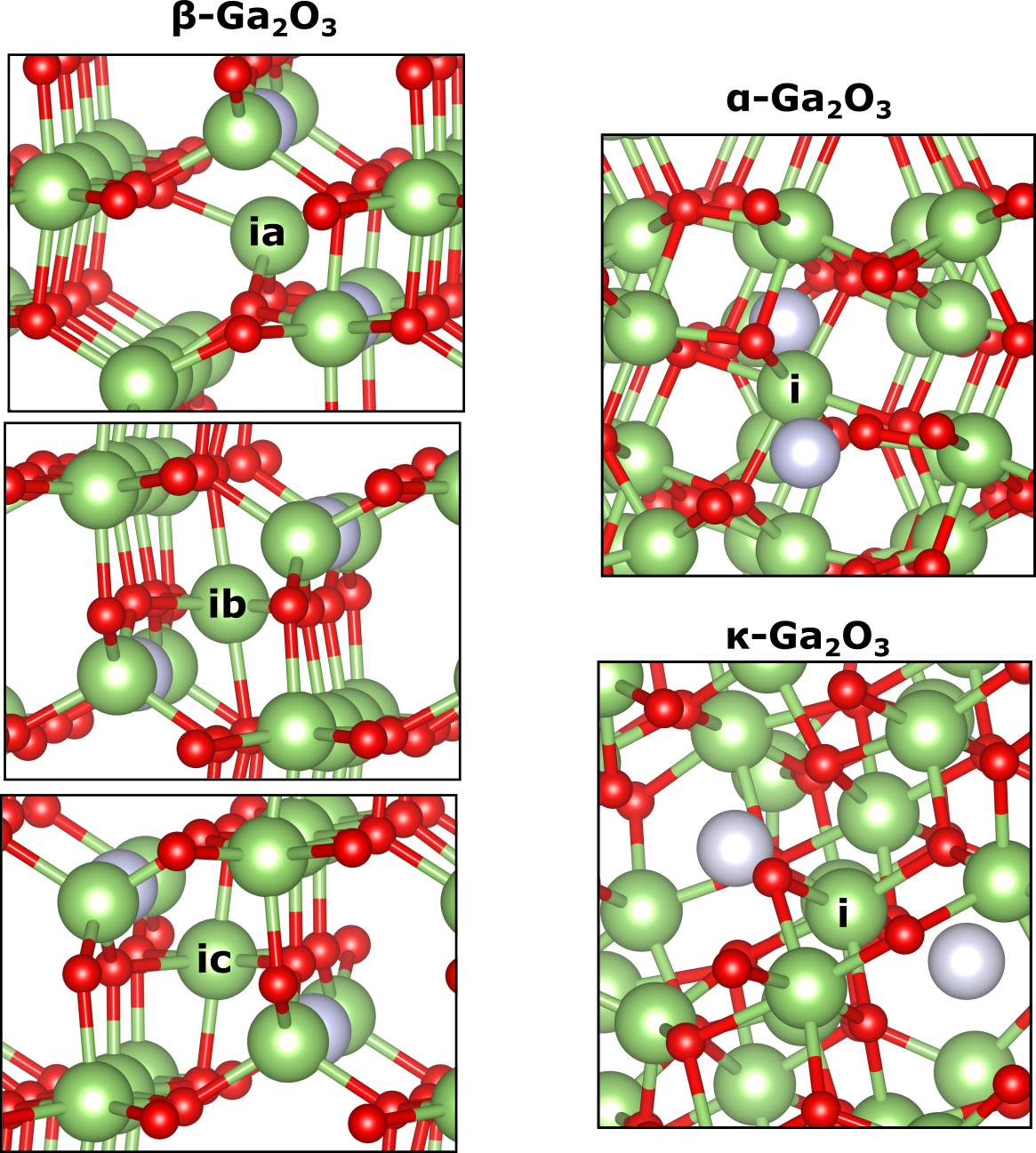} 
     \caption{Relaxed defect structures of split Ga vacancies in $\beta$-, $\alpha$-, and $\kappa$-Ga$_2$O$_3$ polymorphs. Ga ions are represented by green spheres and O atoms by red spheres. Interstitial Ga ions, labelled by ``i'', are surrounded by two Ga vacancies, represented by light grey spheres. In the case of $\beta$-Ga$_2$O$_3$, there are three inequivalent interstitial sites, labelled by ``ia'', ``ib'', and ``ic''.}
     \label{fig:split}
\end{figure}

Split vacancies in $\beta$-\ce{Ga2O3} were first predicted by Varley \emph{et al.} while exploring migration barriers for the Ga vacancies.\cite{varley_hydrogenated_2011}  From our calculations, we accurately identified the previously reported split vacancies ($V_{\text{Ga}}^{\text{ia}}$, $V_{\text{Ga}}^{\text{ib}}$, and $V_{\text{Ga}}^{\text{ic}}$).\cite{tuomisto2023ga}  The split-vacancy $V_{\text{Ga}}^{\text{ic}}$ is strongly preferred over the conventional vacancy V$_{\text{Ga}}$, with $V_{\text{Ga}}^{\text{ic}}$ and $V_{\text{Ga}}^{\text{ib}}$ split vacancies having lower energy by $\sim 0.75$~eV and $\sim 0.20$~eV respectively, which significantly increases their concentrations. Relaxed defect structures of split Ga vacancies in $\beta$-, $\alpha$-, and $\kappa$-Ga$_2$O$_3$ polymorphs are shown in the Figure~\ref{fig:split}.

In $\alpha$-\ce{Ga2O3}, our results show that the split Ga vacancy is $\sim 0.41$~eV energetically lower than the simple vacancy configuration. The energy difference is smaller than in the case of $\beta$-\ce{Ga2O3}.  This trend aligns with recent investigations of other metal oxides isostructural to corundum. \cite{fowler2024metastable,kavanagh_identifying_2025} Previous reports suggested that $\alpha$-\ce{Ga2O3} lacks Ga split-vacancies that characterise $\beta$-\ce{Ga2O3}.\cite{polyakov2022point} Our results confirm that a split-vacancy configuration does exist as an energetically favourable state.  We should note that computational identification of these split-vacancies requires overcoming significant challenges, as the necessary structural change (non-local reconstruction) occurs during the transformation from isolated vacancies to split vacancies. Conventional semi-local structure-searching methods that include only the first few neighbour shells can fail to identify these defect complexes.

A split-vacancy configuration is also stable in $\kappa$-\ce{Ga2O3} phase and more energetically favoured over all single vacancies ($V_{\text{Ga}_1}$-$V_{\text{Ga}_4}$). This partially agrees with a recent report, which also identified a stable $V_{\text{Ga}}^{\text{i}}$ configuration,  although its formation energy was found to lie between $V_{\text{Ga}_1}$ and $V_{\text{Ga}_2}$.\cite{mazzolini2024engineering} In contrast, our calculations show that the split vacancy is in fact the thermodynamic ground-state Ga vacancy configuration with slightly lower energy than the simple vacancy by  $\sim 0.14$~eV. Although this stabilisation energy is much lower than that obtained in $\alpha$-\ce{Ga2O3} and $\beta$-\ce{Ga2O3} phases, substantially larger energy differences, \emph{e.g.}, $\sim 1$~eV, are obtained relative to the other single vacancy configurations.

Importantly, all MLIP relaxations correctly identified the split-vacancy ground state configuration and reproduced the same energy ordering of symmetry-inequivalent split and single vacancy configurations. This confirms the stabilisation relative to the lowest-energy single vacancy across all polymorphs, in good agreement with HSE06 DFT calculations. This agreement indicates that the underlying defect physics is captured robustly by transferable foundation models.
%\newpage
\subsection{Split-Vacancy Preference and Doping Limits}

\begin{figure}[h]
     \centering
\includegraphics[width=17cm]{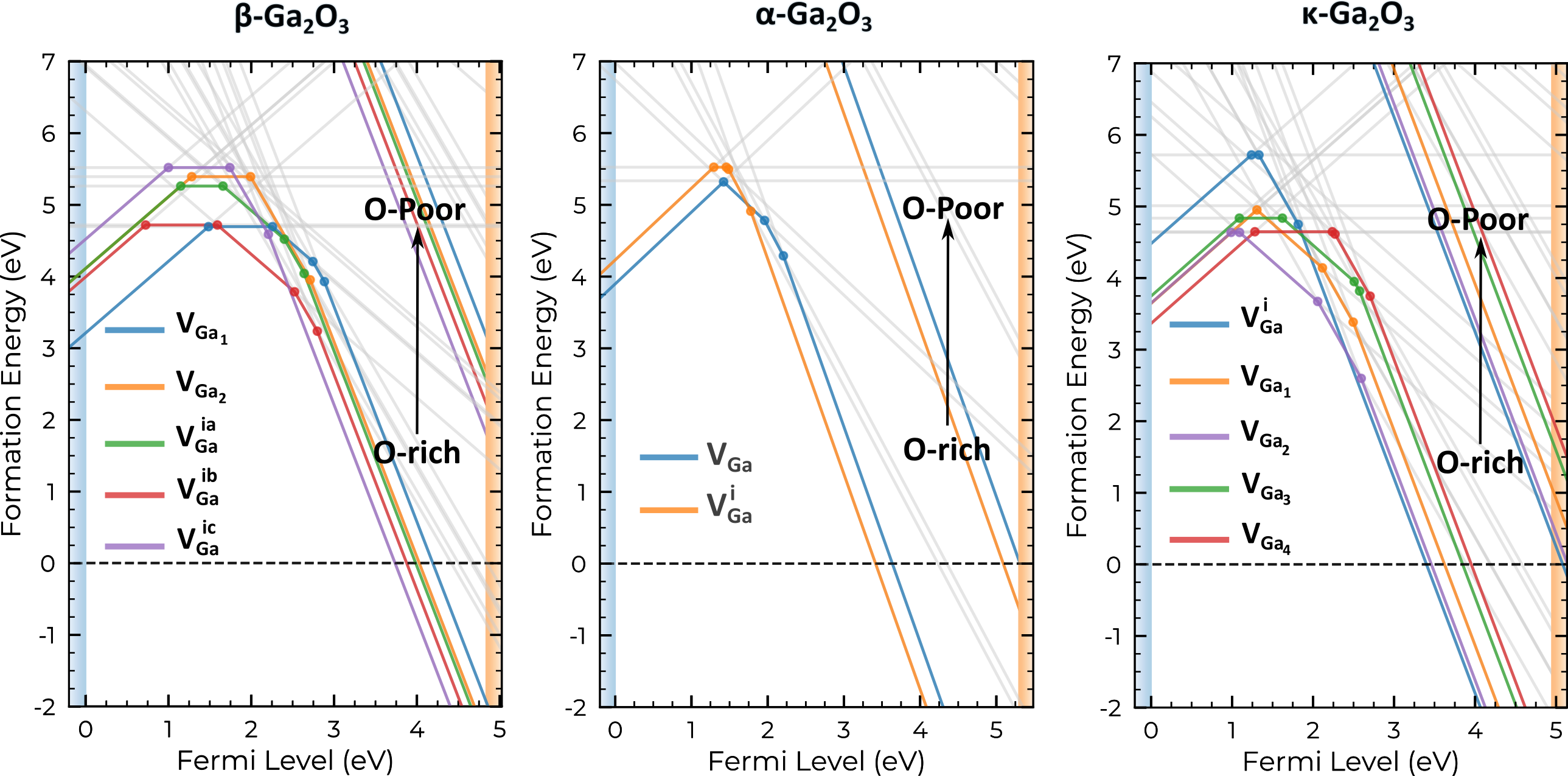} 
     \caption{Defect formation energies in $\beta$-\ce{Ga2O3} , $\alpha$-\ce{Ga2O3} and $\kappa$-\ce{Ga2O3} as a function of the Fermi level for extreme O-rich (Ga-poor) and O-poor (Ga-rich) conditions. The faded lines indicate regions where the defect charge states are unstable.}
     \label{fig-1}
\end{figure}

To further investigate the defect formation across the Fermi level range, we computed the defect formation energies of single and split vacancy configurations for all relevant charge states (shown in Figure~\ref{fig-1}). We considered both oxygen-poor and oxygen-rich conditions -- relative to reference states of Ga and O elements. Our results demonstrate that the split-vacancy is stable and more energetically preferred for all studied polymorphs over single vacancies (\emph{i.e.}, the dominant native acceptor) at a wide range of E$_F$ level positions, especially at higher E$_F$ typical of an n-type material.

As previously discussed, this energy gain is particularly pronounced in the $\beta$-phase. The transition levels for the split vacancies are generally ``deep'', meaning these defects act as potent compensating centres rather than shallow acceptors in all studied polymorphs. Notably, the split vacancy energy remains lower than the single vacancy, even as $E_F$ moves toward the middle of the band gap ($\sim$ [2-3~eV] above valence band maximum). 

Under oxygen-rich conditions, the formation energy of $V_{\text{Ga}}$ decreases significantly, making them more likely to form spontaneously. This suggests that in $n$-type $\text{Ga}_2\text{O}_3$ grown under high oxygen partial pressures, split-vacancy cation vacancies will act as the primary compensation mechanism, limiting the maximum achievable free-electron concentration. Interestingly, while the split-vacancy preference is universal, the doping limits from oxygen-rich to oxygen-poor vary. The $\alpha$-phase shows a slightly lower doping limit compared to $\beta$, which might imply a lower tendency for $n$-type doping even at oxygen-poor conditions. The $\kappa$-phase shows a similar trend, confirming that the ``split" nature of the cation vacancy is an intrinsic property of the Ga-O bonding environment regardless of the specific crystal symmetry.

\begin{figure}[h]
     \centering
\includegraphics[width=17cm]{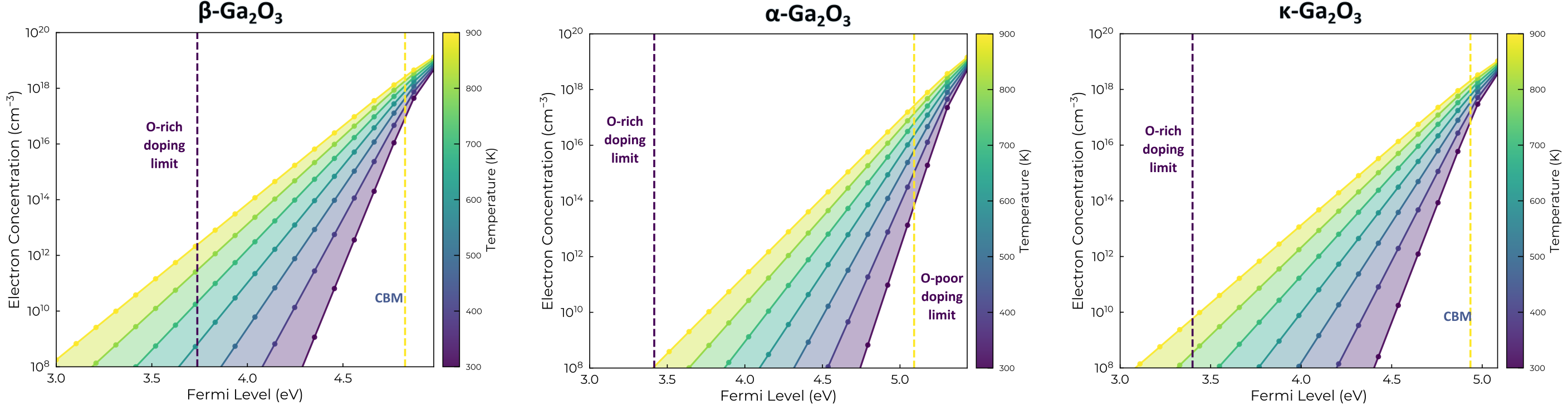} 
     \caption{Electron concentration in $\beta$-\ce{Ga2O3} , $\alpha$-\ce{Ga2O3} and $\kappa$-\ce{Ga2O3} as a function of temperature and Fermi level positions, including extreme O-rich (Ga-poor) and O-poor (Ga-rich) conditions.}
     \label{fig:concentration}
\end{figure}

Under O-rich conditions, the maximum achievable Fermi level lies at 3.74~eV, 3.42~eV and 3.4~eV above VBM  for $\beta$-, $\alpha$-, and $\kappa$-Ga$_2$O$_3$ phases, respectively. Under O-poor conditions, the Fermi level can reach the CBM in both $\beta$- and $\kappa$-phases, while it is limited to 0.19~eV below the CBM in the $\alpha$-phase. To quantify the practical impact of these doping limits, we calculated the electron concentration as a function of Fermi level position and temperature for the three polymorphs (Figure \ref{fig:concentration}). The resulting free electron concentration remains relatively low under O-rich conditions. In contrast, substantially higher carrier concentrations (often exceeding $10^{18}$) become accessible when the Fermi level approaches the CBM under O-poor conditions. Those values give an upper limit of the n-doping of the $\beta$-, $\alpha$-, and $\kappa$-Ga$_2$O$_3$ phases, as the Fermi level can be lower.

\subsection{Si and Hf substitutional donors}

Native acceptor defects such as $V_{\text{Ga}}^{\text{i}}$ can dramatically increase compensation, limiting electron dopability in n-type \ch{Ga2O3} for all polymorphs. To investigate the limits of n-type dopability in detail, we considered two effective dopants, the group-IV element Si and the transition metal Hf. Group-IV elements and transition metals dopants have been widely investigated as effective n-type dopants in $\beta$-\ch{Ga2O3}. Transition metal donors are considered to overcome some of the group IV element limitations, such as the formation of secondary phases at large Si 
concentrations.  

Our calculations indicate a clear site preference that correlates with the local coordination environment of each polymorph.\cite{fadla2025tailoring} Si preferentially occupies tetrahedral site consistent with its smaller ionic radius and Hf preferentially occupies the octahedral site, benefiting from the larger available volume.
\begin{figure}[ht]
     \centering
\includegraphics[width=17cm]{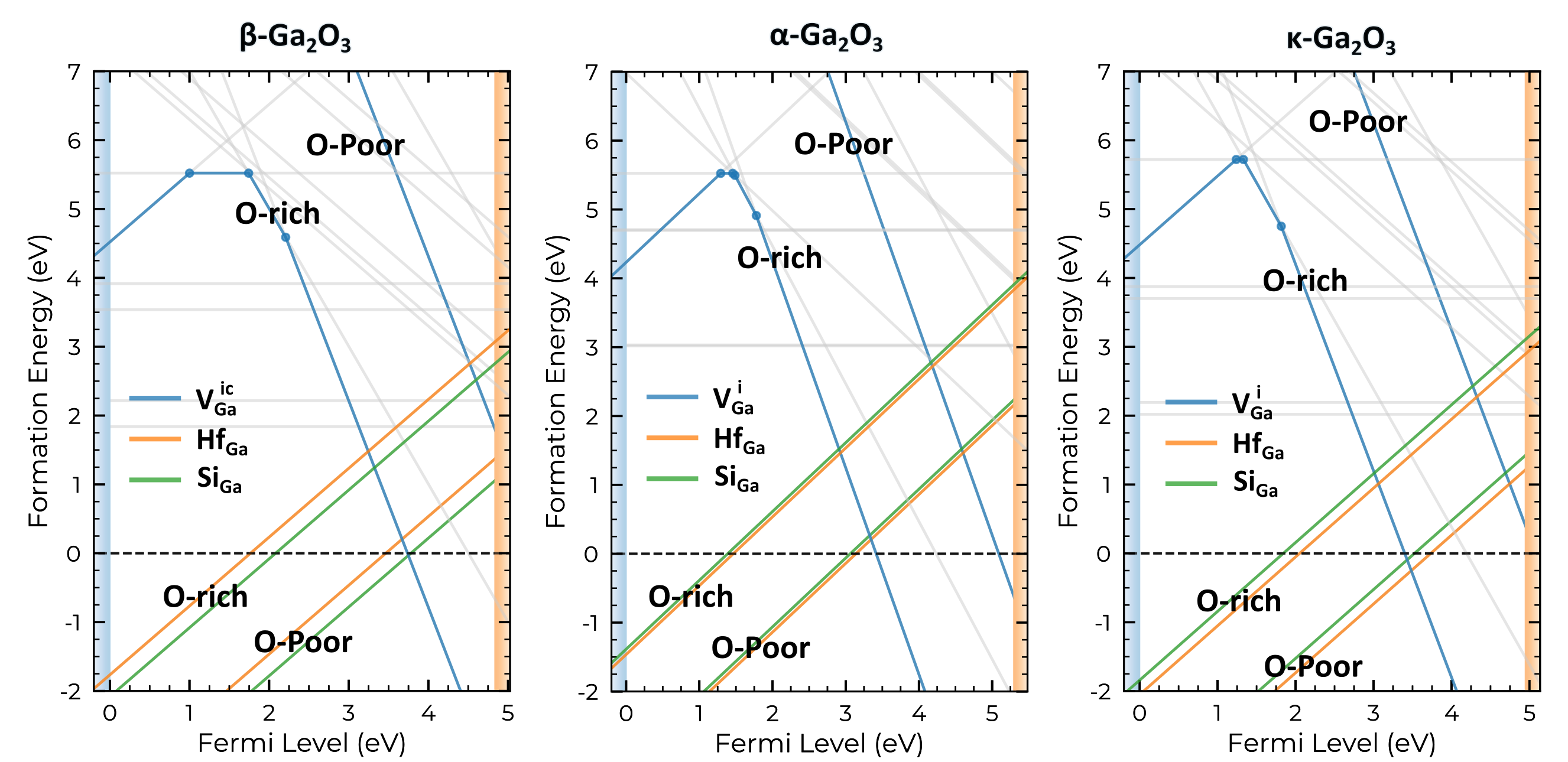} 
     \caption{Defect formation energies for $\mathrm{Hf}_\mathrm{Ga}$, $\mathrm{Si}_\mathrm{Ga}$ substitutional donors and the lowest-energy split-vacancy configuration $V_{\text{Ga}}^{\text{i}}$ in $\beta$-\ce{Ga2O3}, $\alpha$-\ce{Ga2O3}, and $\kappa$-\ce{Ga2O3} as a function of the Fermi level under extreme O-rich (Ga-poor) and O-poor (Ga-rich) conditions. The faded lines indicate regions where the defect charge states are unstable.}
     \label{fig-02}
\end{figure}

As shown in Figure \ref{fig-02}, Si dopants generally display a lower formation energy compared to Hf dopants in  $\beta$-\ce{Ga2O3}. This trend is slightly reversed in the orthorhombic $\kappa$-\ce{Ga2O3}, where Si dopant has a slightly higher formation energy. In the case of the corundum $\alpha$-\ce{Ga2O3}, which consists purely of octahedral sites, both dopants exhibit nearly identical formation energies. 

Under extreme oxygen-poor conditions, both Si and Hf dopants maintain significantly lower formation energy compared to the split vacancy, which in the case of $\beta$-\ce{Ga2O3}, across the entire Fermi level range up to the CBM. This indicates that high doping efficiency is attainable under oxygen-poor growth conditions. However, under oxygen-rich (Ga-poor) conditions, the formation energy of the triple-charged split drops precipitously, thus preventing the material from becoming degenerately n-type. All dopant transition levels in \ce{Ga2O3} polymorphs occur above the CBM. Thus, both Si and Hf act as shallow donors.

Our results suggest that growth under oxygen-poor conditions is critical to suppressing the self-compensation and maximising free-electron concentrations, particularly in the $\alpha$- and $\kappa$-phases where the compensation onset occurs at lower Fermi levels.

\section{Conclusions}
In this work, we employed a machine learning foundation model to accelerate the discovery of split-vacancy configurations, followed by high-level atomistic modelling using the Heyd-Scuseria-Ernzerhof (HSE06) hybrid functional level. In particular, single vacancy $V_{\text{Ga}}$, split vacancy $V_{\text{Ga}}^{\text{i}}$ and substitutional donors ($\mathrm{Hf_{Ga}}$ and $\mathrm{Si_{Ga}}$)  were investigated to determine their impact on n-type dopability in \ce{Ga2O3} polymorphs.

Our results reveal that the split vacancies are stable and preferable across $\beta$-, $\alpha$-, and $\kappa$-\ce{Ga2O3} polymorphs. This confirms that the split vacancy is not a $\beta$-specific phenomenon, but an intrinsic feature of Ga$-$O bonding. Split vacancy configurations act as deep acceptors with low formation energy across a wide Fermi-level range, representing the primary self-compensation mechanism in n-type material across all phases. Notably, the $\alpha$-phase exhibits a lower compensation threshold than $\beta$-phase, suggesting it is challenging to achieve heavy n-doping.

Hf and Si show low formation energy and act as shallow donors, especially under oxygen-poor conditions, with Si favouring tetrahedral sites and Hf octahedral sites. Because split vacancies are least stable under oxygen-poor conditions, maintaining oxygen-poor growth conditions is critical for suppressing compensation and maximising free-electron concentration in all \ce{Ga2O3} polymorphs.

MLIPs have proven highly effective for rapid screening and pre-relaxation, successfully identifying the non-local reconstructions that traditional local structure searches often miss. Beyond accelerating structure searches, both transferable foundation models and HSE06-trained MLIPs successfully captured the relative energetics of complex non-local split-vacancy reconstructions across all \ce{Ga2O3} polymorphs. MLIPs yielded the same energy ranking as HSE06 calculations, highlighting the emerging capability of modern interatomic potentials for predictive defect physics.

\newpage

%\section{Associated content}

\begin{acknowledgements}
This work was supported by a research grant from the Department for the Economy Northern Ireland (DfE) under the US-Ireland R\&D Partnership Programme (USI 195).
Access to the computing facilities and support from the Northern Ireland High-Performance Computing (NI-HPC) service funded by EPSRC (EP/T022175), and to the UK national high-performance computing service, ARCHER2, through the UKCP consortium and funded by EPSRC (EP/X035891/1) are also gratefully acknowledged.

\end{acknowledgements}
\section*{Data Availability Statement}
The data that support the findings of this study are openly available at [link] (released upon publication)

\bibliography{reference}

\end{document}